\documentclass[aip,pre,twocolumn,superscriptaddress,reprint]{revtex4-2}

\usepackage{amsfonts}
\usepackage{amsmath}
\usepackage{amssymb}
\usepackage{graphicx}
\usepackage{float}
\usepackage[dvipsnames]{xcolor}
\begin{document}

\title{A simple time coarse graining method for molecular dynamics simulations of liquids.}


\author{Maxime Martin}
\affiliation{Laboratoire de Photonique d'Angers EA 4464, Universit\' e d'Angers, Physics Department,  2 Bd Lavoisier, 49045 Angers, France.}

\author{Levi Pereon}
\affiliation{Laboratoire de Photonique d'Angers EA 4464, Universit\' e d'Angers, Physics Department,  2 Bd Lavoisier, 49045 Angers, France.}

\author{Quoc Tuan Truong}
\affiliation{Laboratoire de Photonique d'Angers EA 4464, Universit\' e d'Angers, Physics Department,  2 Bd Lavoisier, 49045 Angers, France.}

\author{Victor Teboul}
\email{victor.teboul@univ-angers.fr}
\affiliation{Laboratoire de Photonique d'Angers EA 4464, Universit\' e d'Angers, Physics Department,  2 Bd Lavoisier, 49045 Angers, France.}


\begin{abstract}
Considering molecular dynamic simulations as a stochastic method, we investigate the possibility of time coarse graining the simulations.
Similarly to Boltzmann inversion method in spatial coarse graining, which begins with a free energy called potential of mean force $V^{mf}(r)=-kT log (g(r))$, we test the effect of a generalized potential of mean force that uses the distinct part of the Van Hove correlation function with a characteristic time different from zero $V^{gmf}(r,\Delta t)=-kT log (G_{d}(r,\Delta t)/\rho)$. We show that the method is approximately equivalent to replace the hard core of the original potential by a smooth harmonic function. We then compare the results of simulations using the modified potential and the original one. 
Results show that this simple modification of the potential, namely replacing the short range wall with a smooth quadratic law, leads to a shift in the time step resulting in {\color{black} a similar dynamics as} the original potential function but with a much larger time step.

\end{abstract}

\maketitle
{
\section{ Introduction}

Molecular dynamics (MD) and Monte Carlo (MC) simulations\cite{md1,md2,md2b} are important tools together with theories and experiment to unravel unsolved problems in condensed matter physics\cite{ms1,ms5,ms6,rate,ms2,silica,argon,ms3,ms4}, chemistry and biology\cite{bio1,bio2,bio3,bio4,bio5}. While computers are becoming faster each year, the main limit in Molecular dynamics simulations nevertheless still remains the limited simulation time. 
This is particularly the case when studying biological systems, or dynamics in viscous media, as for example liquids in their approach to the glass transition\cite{gt0,gt1,gt2,anderson,fragile1}.
To solve that problem, one may coarse grain the potential, use parallelization, or simply increase the time step of the simulation.
However when increasing the time step in MD, molecules at some point begin to overlap, leading to infinite energies due to the short range wall of intermolecular potentials and thus breaking the simulation.\\
A simple method to decrease the probability of infinite energies is to trim and smooth partly the potential wall.
We may expect that as the wall is scarcely probed by the simulations due to its high energies, the results will be only slightly affected by the smoothing.\\
Another method is coarse graining.
Coarse graining methods\cite{coarse1,coarse2,coarse3,coarse4,coarse5,coarse6,coarse7} are often used to simplify the molecules and accelerate the simulations. 
The most often used Boltzmann inversion method, begins with the potential of mean forces\cite{md1} $V^{mf}=-kT  log(g(r)) $ (in fact a free energy) that comes from the radial distribution function $g( r) $ and Boltzmann probability distribution. 

In this work we raise the question of using a similar method for time coarse graining, and its connection with the smoothing of the potential wall. 
It has long been emphasized that because the time step is finite in molecular dynamics, the simulated trajectories are not exact,
but escape the Newtonian trajectories after a few thousands steps, a number of step that depends on Lyapunov exponent. 
As a result MD is not actually determinist but can be seen as a stochastic method, opening the possibility of time coarse graining methods.

Following Boltzmann inversion method, we use a generalization  of the potential of mean force using the distinct Van Hove correlation function $G_{d}(r,\Delta t)/\rho$, {\color{black} where $\rho=N/V$ is the number density,} instead of the radial distribution function $g(r)=G_{d}(r,\Delta t=0)/\rho$, leading to the function:\\ 
 \begin{equation}
\displaystyle{V^{gmf}(\Delta t)=-kT  log(G_{d}(r,\Delta t)/\rho)                           }       \label{e11}     
\end{equation}
{\color{black} that may be used as a first approximation for a time coarse grained potential} in MD using a time step $\Delta t$. 
Interestingly enough we will see that this method is approximately equivalent to a smoothing of the short range wall of the potential with an harmonic function.
{\color{black} Using this result, our coarse grained potential will then be made from a quadratic short range wall, and the Lennard-Jones original potential at larger ranges.
We keep the long range part of the potential in order to not overestimate the interaction with a mean field approach as multiple atomic interactions will be used in the simulations. We attach these two potentials with a continuous derivative, for the force to be continuous, at a distance $R$ that defines the quadratic short-range part and therefore our potential.}\\

{\color{black} Notice however, that if the mean force potential is in fact a free energy,  our generalized mean force potential has no simple static interpretation.
It can be connected to the path probability exponent that replaces the potential function in trajectory space, or to the hopping probability in a dynamic Monte Carlo simulation. 
In this work, we will call it anyway 'generalized mean force potential' (GMF) for simplicity as it has the dimension of an energy.}\\

The created effective GMF potential is intended to be used not only in advanced molecular dynamics simulations but also in Dissipative Particle Dynamics\cite{dpd1,dpd2,dpd3,dpd4,dpd5,dpd6}, and dynamic Monte Carlo simulations. It may also permit to understand the effects of time coarse graining in relation with the glass transition, because the glass transition is in facilitation theories seen as a transition involving time (transition in the trajectory space \cite{facile,facile1,facile2,facile3,facile4,facile5,facile6}). In this viewpoint the time temperature superposition effect\cite{timetemperature00,timetemperature01,timetemperature02,timetemperature03,timetemperature,timetemperature2,timetemperature3,timetemperature4} could be understood as equivalent to a renormalization in time. To test that {\color{black} interpretation} we need a method for time renormalization or equivalently time coarse graining that we implement in this work.

\section{Calculation}

The aim of this work is  to create an effective potential for time coarse graining. For that purpose we begin as in spatial coarse graining with a mean force potential. 
In order to apply the mean force potential to a coarse graining in time, we replace the radial distribution function by the distinct van Hove correlation function $G_{d}(r,\Delta t)$.
That function represents the probability distribution to find an atom at a distance $r$ from the original position of the other atoms after a time lapse $\Delta t$.\\

Our test medium is a minimal model liquid\cite{ariane} chosen to hinder crystallization and accelerate the simulations. It is constituted of dumbbell diatomic molecules, with rigidly bonded atoms (indexed as $\alpha$ or $\beta =1, 2$) with a fixed interatomic distance of $l = 1.73$ Å. The two atoms are defined with the same mass $m_{0}=20g/N_{A}$.    {\color{black} An atom i, of type $\alpha$, interacts with an atom j, of type $\beta$, through the following Lennard-Jones potential:}
\begin{equation}
V^{LJ}_{\alpha \beta }(r_{ij})=4\epsilon_{\alpha \beta}((\sigma_{\alpha \beta}/r_{ij})^{12} -(\sigma_{\alpha \beta}/r_{ij})^{6})   \label{e1}
\end{equation}
The parameters of the potential\cite{ariane} are: $\epsilon_{11}= \epsilon_{12}=0.25 KJ/mol$, $\epsilon_{22}= 0.2 KJ/mol$,  $\sigma_{11}= \sigma_{12}=3.45$\AA, $\sigma_{22}=3.28$\AA.
{\color{black} The force ${{f}}^{LJ}_{\alpha \beta}(r_{ij})$ between atoms i and j is then obtained from the potential derivative ${{f}}^{LJ}_{\alpha \beta}(r_{ij})= - {dV^{LJ}_{\alpha \beta}(r_{ij})\over dr_{ij}}$ and  ${\bf{f}}^{LJ}_{\alpha \beta,ij}(r_{ij})={{f}}^{LJ}_{\alpha \beta}(r_{ij}). {{\bf{r}}_{ij}/ r_{ij}}$.
}
The length of the molecule is $l_{m}=5.09$\AA\ and its width $L_{m}=3.37$\AA.
Our cubic simulation boxes contains $1000$ molecules {\color{black} with  sides of $32.49$ \AA, or $2000$ molecules with sides of $40.9$ \AA. 
We use the Gear algorithm with the quaternion method to solve the equations of motions with a $10^{-15} s$ time step.
A  Berendsen thermostat\cite{berendsen} } removes the energy dissipated into the system avoiding a  drift in energy. 
{\color{black} Consequently, the number of molecules N, volume V and temperature T are fixed in our study.}
The temperature is maintained constant at T=500K, a temperature a hundred degrees below the melting temperature of our medium.
{\color{black} We used $20 ns$ simulation runs beginning with a previously equilibrated simulation box in order to avoid aging.
Notice that due to the use of Lennard Jones potentials only, the potential can be easily shifted to model although approximately a large number of real viscous liquids.}\\

 {\color{black} We show in Figure \ref{fig1} that the logarithm of the distinct Van Hove correlation function can be fitted with a quadratic function, a result that we will detail later.
 We then use this result, to model our generalized coarse grained potential from a quadratic short range wall, and the Lennard-Jones original potential at larger ranges. We attach these two potentials with a continuous derivative, for the force to be continuous, at a distance $R$ that defines the quadratic short-range part and therefore our potential. Doing so, we choose to use he same distance $R$ for every atomic bead, leading then to a slightly different short range potential for each bead.
 The time evolution of the results are then tested for various $R$ values, showing the acceleration of the dynamics. This acceleration can be used directly with the original time step ($10^{-15} s$) as in this work or amplified using a larger time step. We found that due to the decrease of the potential wall, time steps $4$ times larger can be used in most cases.}\\

To test our coarse graining procedure, we will investigate the possible alteration of various correlation functions using the new potentials.
We calculate the diffusion coefficient, mean square displacement, incoherent scattering function, $\alpha$ relaxation time which is associated with the viscosity of the medium, the non Gaussian parameter, and more importantly the self and distinct Van Hove correlation functions. 
Let's now define the statistical functions utilized for this purpose\cite{book}. One function of significant relevance in glass-transition phenomena is the intermediate scattering function, denoted as $F_{S}(Q,t)$, which portrays the autocorrelation of density fluctuations at the wave vector $Q$. This function provides insights into the structural relaxation of the material. We define $F_{S}(Q,t)$ through the following relation:

\begin{equation}
\displaystyle{F_{S}(Q,t)={1\over N N_{t_{0}}} Re( \sum_{i,t_{0}} e^{i{\bf Q.(r_{i}(t+t_{0})-r_{i}(t_{0}))}}  )          }\label{e1}
\end{equation}
For physical reasons, Q is chosen as the wave vector (here $Q_{0}=2.25$\AA$^{-1}$) corresponding to the maximum of the structure factor $S(Q)$.
$F_{S}(Q_{0},t)$ then allows us to calculate the $\alpha$ relaxation time $\tau_{\alpha}$  of the medium from the equation: 
\begin{equation}
\displaystyle{F_{S}(Q_{0},\tau_{\alpha})=e^{-1}}       \label{e10}     
\end{equation}
The diffusion coefficient $D$ is obtained from the long time limit of the mean square displacement $<r^{2}(t)>$:
\begin{equation}
\displaystyle{<r^{2}(t)>=   {1\over N N_{t_{0}}}  \sum_{i,t_{0}}  ({\bf r}_{i}(t+t_{0})-{\bf r}_{i}(t_{0}))^{2}                                }       \label{e11}     
\end{equation}
and
\begin{equation}
\displaystyle{\lim_{t \to \infty}  <r^{2}(t)>=  6 D t                               }       \label{e12}     
\end{equation}

Most importantly we will rely on the Self and distinct part of the Van Hove correlation functions\cite{book} to describe the structure and dynamics of the medium.

The self part of the Van Hove correlation function $G_{s}(r,\Delta t)$ represents the probability distribution for a particle to be after a time lapse $\Delta t$ a distance $r$ apart its previous position. The function describes the statistical displacements of the molecules.

\begin{equation}
{\color{black} \displaystyle{G_{s}(r,\Delta t)=   {1\over N N_{t_{0}}}  \sum_{i,t_{0}}  \delta(r-\lvert{\bf r}_{i}(t_{0}+\Delta t)-{\bf r}_{i}(t_{0})\rvert)                                }   }    \label{e11a}     
\end{equation}

The distinct part of the Van Hove correlation function\cite{book} $G_{d}(r,\Delta t)$ represents the probability distribution for a particle to be after a time lapse $\Delta t$ a distance $r$ apart the previous position of its surroundings. It describes statistically the dynamics of the structure of the medium.

\begin{equation}
{\color{black}\displaystyle{G_{d}(r,\Delta t)=   {1\over N N_{t_{0}}}  \sum_{i, j \neq i,t_{0}}  \delta(r-\lvert{\bf r}_{i}(t_{0}+\Delta t)-{\bf r}_{j}(t_{0})\rvert)                                }}       \label{e11b}     
\end{equation}

The radial distribution function $g(r)$  is directly related to the distinct part of the Van Hove for a zero time lapse:
\begin{equation}
\displaystyle{g(r) = G_{d}(r,0)/\rho                             }       \label{e11c}     
\end{equation}

Finally, we also use the non Gaussian parameter $\alpha_{2} (t)$ that quantifies the deviation of $G_{s}(r,\Delta t)$ from the Gaussian form predicted by Brownian motion. In supercooled liquids $\alpha_{2} (t)$ measures cooperative motions, as they induce a tail in the Van Hove.

\begin{equation}
\displaystyle{\alpha_{2}(t)=   {3 <r^{4}(t)>\over 5<r^{2}(t)>^{2} } -1                               }       \label{e11d}     
\end{equation}


\section{Results and discussion}

\begin{figure}
\centering
\includegraphics[height=7.5 cm]{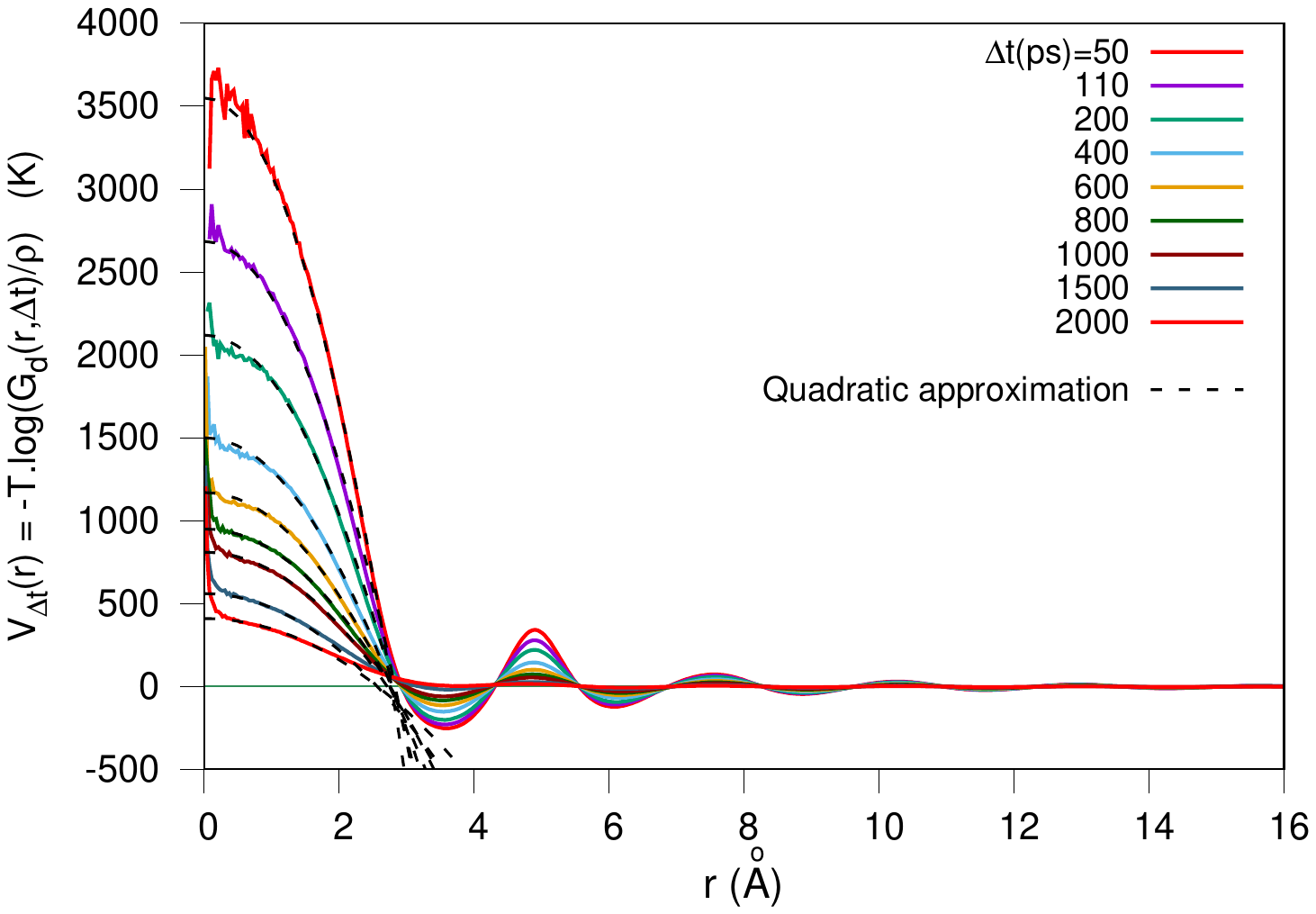}

\caption{(color online)  Generalized mean force approximation using the distinct part of the Van Hove correlation function with a characteristic time $\Delta t$,  $V^{gmf}(r,\Delta t)=-kT log (G_{d}(r,\Delta t)/\rho)$ (lines), compared with a quadratic law of the form {\color{black} $V_{g}(r,\Delta t)=a_{\Delta t} - b_{\Delta t}.r^{2}$ (dashed lines).
The quadratic law fits well the $V^{gmf}(r,\Delta t)$ potential.
Using this approximation, our effective potential is then created from a quadratic law at short range, followed by the Lennard-Jones potential at larger ranges.
The long range interactions are not modified in order not to overestimate multiple interactions in simulations.
The two functions join at a distance $R$ with a continuous slope insuring that the force is continuous. } 
}
\label{fig1}
\end{figure}

We display in Figure \ref{fig1} the generalized mean force potential $V^{gmf}(r,\Delta t)=-kT log(G_d(r,\Delta t)/\rho)$ where the distinct Van Hove correlation function   on a time step $\Delta t$ replaces the radial distribution function $g(r)$.
As $\Delta t$ increases, the height of the short range wall decreases, and the following oscillations are smoothed down.
Interestingly the short range wall of the {\color{black} GMF potential} follows a quadratic law of the form: \\

{\color{black}
$V_{g,\alpha \beta}(r_{ij},\Delta t)=a_{\alpha \beta,\Delta t} - b_{\alpha \beta,\Delta t}.r_{ij}^{2}$
 
where  $a_{\alpha \beta,\Delta t}$ and $b_{\alpha \beta,\Delta t}$ are fit parameters that depend on $\Delta t$ and Lennard-Jones atomic interactions, and we specified the parameter $r$ as the interatomic distance $r_{ij}$.}\\

{\color{black}The quadratic law may result here from the motion of the molecule inside the cage, as interaction potentials are approximately quadratic around an equilibrium position in a Taylor expansion.}
This result suggests that a simple modification of the original intermolecular potential $V_{\alpha \beta}^{LJ}(r)$, namely replacing the short range wall with a smooth quadratic law, could lead to a shift in the time step resulting in {\color{black} a similar dynamics} as the original potential function $V_{\alpha \beta}^{LJ}(r)$ but with a much larger time step.
We will now test this very simple time coarse grained modeling and show that it leads approximately to the same dynamics as the original potential but with a much larger time step. 
For that purpose we will study the possible modifications of the radial distribution function $g(r)$, the self and distinct part of the Van Hove correlation functions $G(r,\Delta t)$, the mean square displacement $<r^{2}(t)>$, the non Gaussian parameter $\alpha_{2}(t)$, the incoherent scattering function $F_{s}(Q,t)$ and of the $\alpha$ relaxation time $\tau_{\alpha}$. 

In these studies for the sake of simplicity, we will replace the {\color{black} set of parameters $a_{\alpha \beta,\Delta t}$ and $b_{\alpha \beta,\Delta t}$,} by a single parameter $R$ that will define our coarse grained potential.  We define $R$ as the cutoff radius below which our potential has been replaced by the quadratic law {\color{black}$V_{g,\alpha \beta}(r_{ij},\Delta t)=a_{\alpha \beta,\Delta t} - b_{\alpha \beta,\Delta t}.r_{ij}^{2}$.}  Enforcing the continuity of the  potential and of its gradient (then the force) at the cutoff value $R$, then results in {\color{black} a single set of parameters $a_{\alpha \beta,\Delta t}$ and $b_{\alpha \beta,\Delta t}$ for each value of $R$. 
More explicitely $a_{\alpha \beta,\Delta t}$ and $b_{\alpha \beta,\Delta t}$ are related to $R$ with the equations:\\
 \begin{equation}
\displaystyle{a_{\alpha \beta,\Delta t}=4\epsilon_{\alpha \beta}(7(\sigma_{\alpha \beta}/R)^{12}-4(\sigma_{\alpha \beta}/R)^{6})}\label{e1}
\end{equation}
\begin{equation}
\displaystyle{b_{\alpha \beta,\Delta t}=12{\epsilon_{\alpha \beta}\over R^2} (2(\sigma_{\alpha \beta}/R)^{12}-(\sigma_{\alpha \beta}/R)^{6})}\label{e1}
\end{equation}

But using the force instead of the potential, leads to more simple relations:\\
 {\color{black} 
\begin{equation}
r\leq R :   {\bf{f}}_{\alpha \beta,ij}(r_{ij})= f_{\alpha \beta}^{LJ}(R).{\bf{r}}_{ij}/R
\end{equation}
\begin{equation}
r > R:  {\bf{f}}_{\alpha \beta,ij}(r_{ij})= {\bf{f}}^{LJ}_{\alpha \beta,ij}(r_{ij})=f_{\alpha \beta}^{LJ}(r_{ij}).{\bf{r}}_{ij}/r_{ij}
\end{equation}
}}\\

 {\color{black} We may also notice the relation: $f_{\alpha \beta}^{LJ}(R)=2b_{\alpha \beta,\Delta t}.R$\\}

\begin{figure}
\centering
\includegraphics[height=7.5 cm]{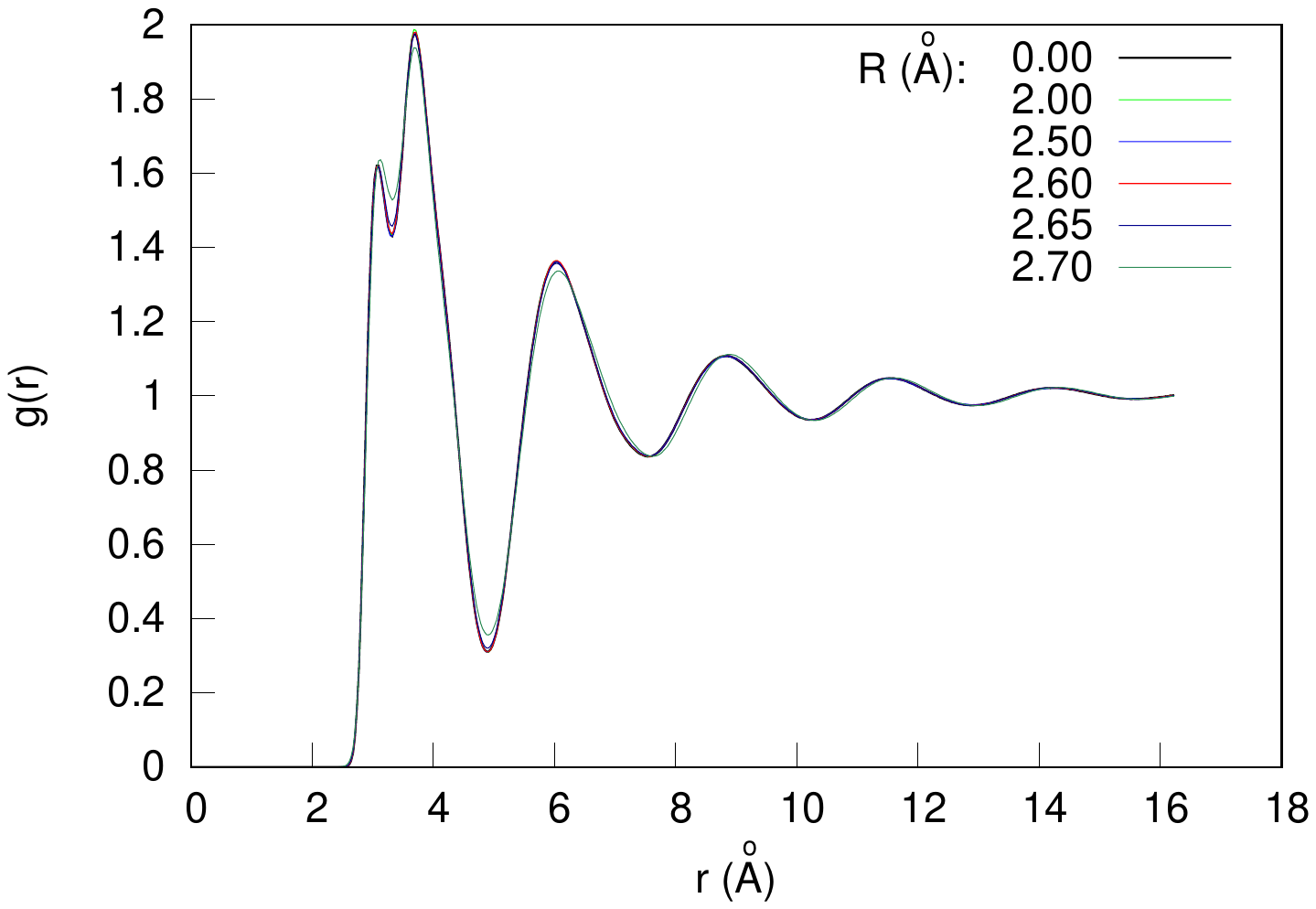}

\caption{(color online)  Radial distribution function for various cutoff radii $R \leq 2.70$ \AA, compared to the original value ($R = 0$ \AA).}
\label{fig2a}
\end{figure}

\begin{figure}
\centering
\includegraphics[height=7.5 cm]{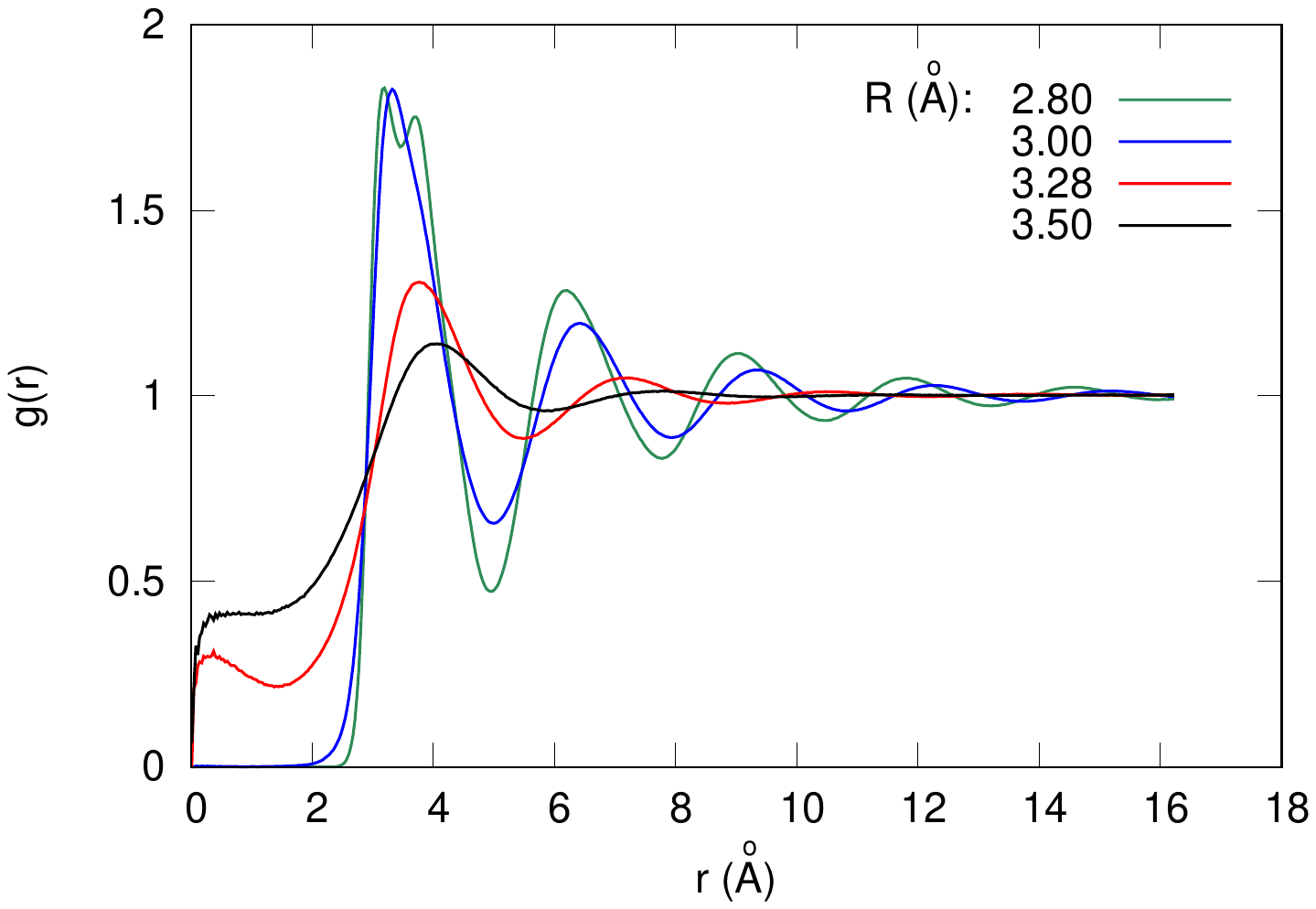}

\caption{(color online)  Radial distribution function for various cutoff radii $R>2.70$ \AA.
}
\label{fig2b}
\end{figure}

Figure \ref{fig2a} shows that below a threshold value $R_{T}=2.65$\AA\ the radial distribution function $g(r)$ (RDF) is unchanged when using the effective potential instead of the real one. 
We interpret that result as follows. As $R$ increases, the height of the wall decreases leading at some point to an energy that can be reached at the temperature of study (T=500K). Below that value the statistics of molecules probing the wall is very scarce leading to no modification of the RDF.
{\color{black} Due to Boltzmann statistics we thus expect $R$ to depend on the temperature and on the Lennard-Jones interaction only.}

Then above the threshold (Figure \ref{fig2b}) we observe a smoothing of the RDF $g(r)$, while the peaks decrease in height and the probability for a molecule to be at short distance from its neighbor, that is inside the original potential wall, increases.  
The RDF $g(r)$ then resembles the distinct part of the Van Hove correlation function $G_{d}(r,\Delta t)/\rho$ for a finite (non zero) time difference $\Delta t$. 
As $R$ increases, $g(r)$ resembles  $G_{d}(r,\Delta t)/\rho$ with $\Delta t$ increasing (see Figure \ref{fig4}).
 
\begin{figure}
\centering
\includegraphics[height=7.5 cm]{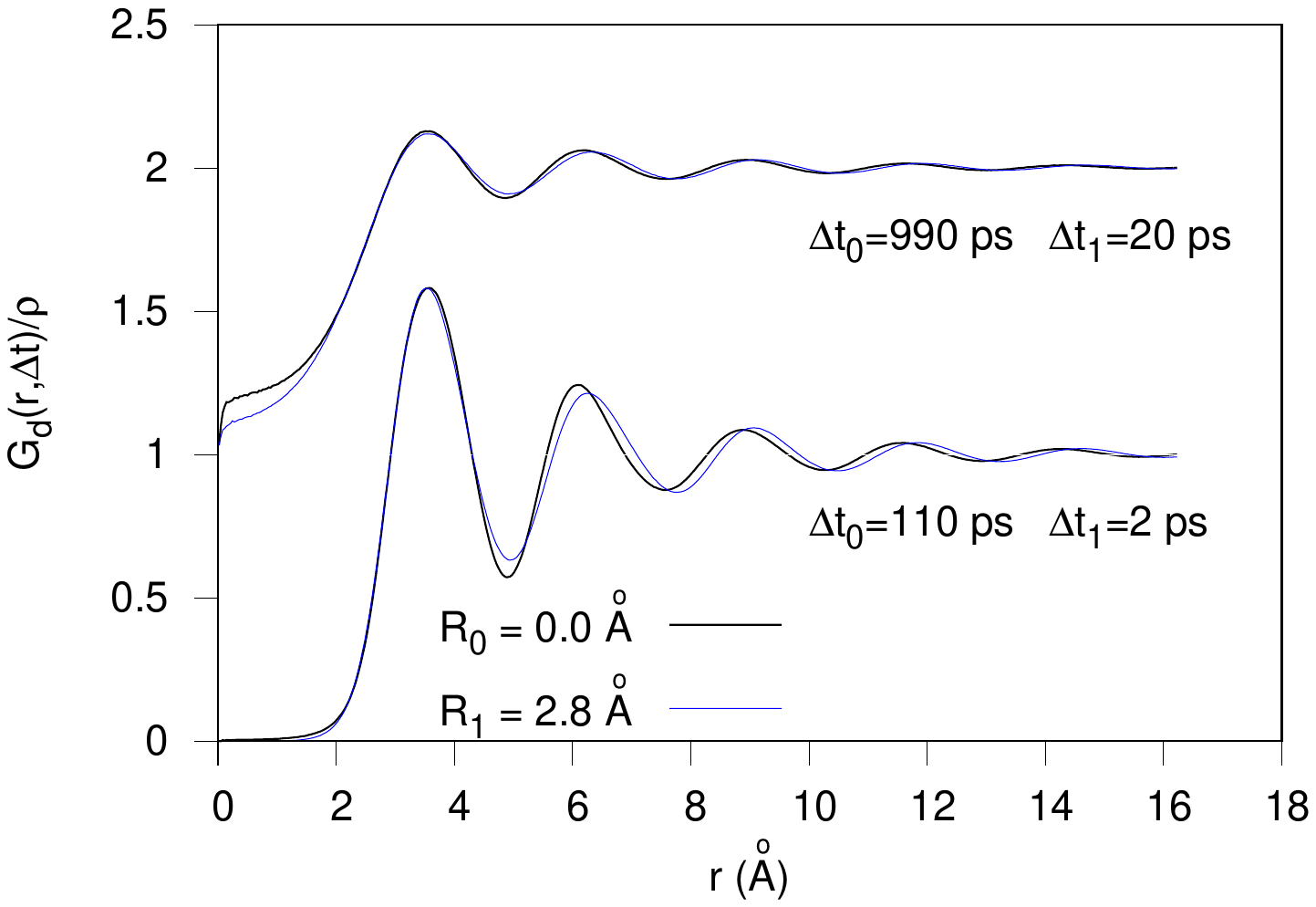}

\caption{(color online)  Distinct part of the Van Hove correlation function $G_{d}(r,\Delta t)$ for $R=2.8$ \AA\  compared with the original potential distinct Van Hove function ($R=0$ \AA). We observe a time shift between the two functions as expected for a coarse graining in time. With the time shift the correlation function is relatively well reproduced.}
\label{fig4}
\end{figure}

Notice that the possibility to find a molecule inside the original potential wall shows that previously forbidden moves do not break the simulation anymore. This is necessary because as we increase the time step, the probability of these moves increases. 
Snapshots (not displayed) do not show any visible modifications of the structure that appears amorphous as in the liquid state.\\


We will now focus on dynamical properties.
In MD simulations the structural calculations usually do not need much effort, while probing the dynamics usually need long time simulations when the medium is viscous. A time coarse grained model is then mostly needed for dynamical data. 
To investigate possible differences between the coarse grained and the original dynamics, we begin with the Van Hove correlation functions which contain the whole dynamical statistical information. Then we will turn to more practical autocorrelation functions like the mean square displacement.\\

Figure \ref{fig4} compares the distinct part of the Van Hove correlation function for the coarse grain potential simulation with $R=2.8$\AA\ and the original potential simulation ($R=0$ \AA\ ), {\color{black} while figure \ref{fig4b} shows the time ratio evolution of the fit between the two distinct Van Hove correlation functions.}   We have chosen $R=2.8$\AA\  to be larger than the threshold value while not too much so that the coarse graining effects are still {\color{black} reasonable.} We found actually that the coarse graining effect increases quite rapidly above the threshold. 
We observe in Figure \ref{fig4} that the coarse grain simulation reproduces almost exactly the distinct part of the Van Hove correlation function $G_{d}(r,\Delta t)$, but with a time difference approximately $50$ times shorter. 
In other words we find the same dynamics but with a much shorter simulation time.
This result  leads to an acceleration of simulations here by a factor $50$ obtained with a simple smoothing of the potential wall, and comes in support of a time coarse graining.
Notice however that while the curves also superimpose themselves in the second plot, the time factor between the coarse grained simulated Van Hove and the original one is not exactly the same for the two plots. A factor $9$ for $\Delta t_{0}$ is converted into a factor $10$ for $\Delta t_{1}$. Therefore the time ratio between the two simulations  $\Delta t_{0}/\Delta t_{1}$ {\color{black}(Figure \ref{fig4b})} decreases slightly as the correlation times become larger.\\

\begin{figure}
\centering
\includegraphics[height=7.5 cm]{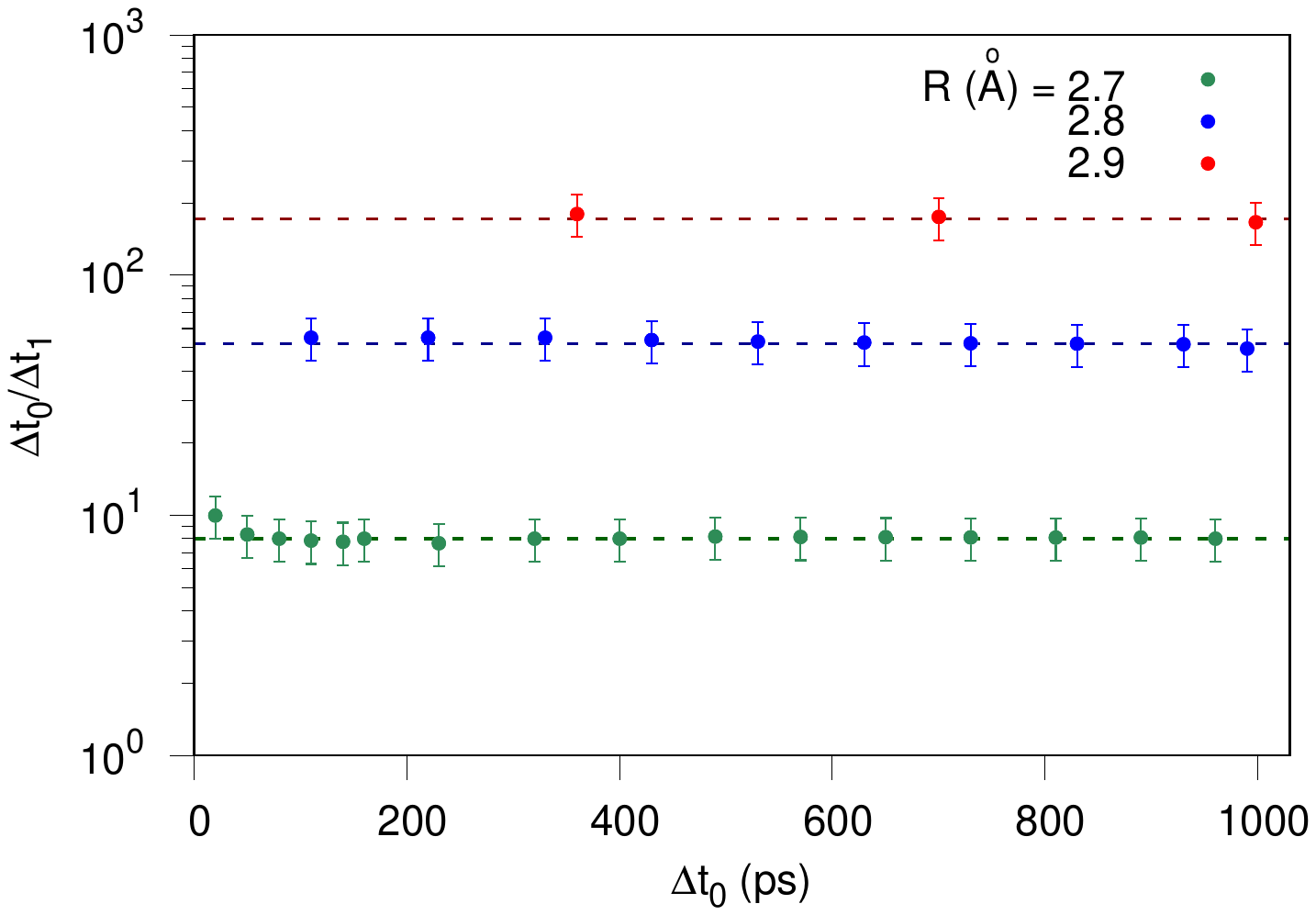}

\caption{(color online)  {\color{black}Time rescaling ratio for the distinct part of the Van Hove correlation function, $\Delta t_{0}/\Delta t_{1}$ versus $\Delta t_{0}$,   where $G_{d}(r,\Delta t_{1})$ fits $G_{d}(r,\Delta t_{0})$, for (a) $R=2.7$ \AA, (b) $R=2.8$\AA, (c) $R=2.9$ \AA.}}
\label{fig4b}
\end{figure}

\begin{figure}
\centering
\includegraphics[height=7.5 cm]{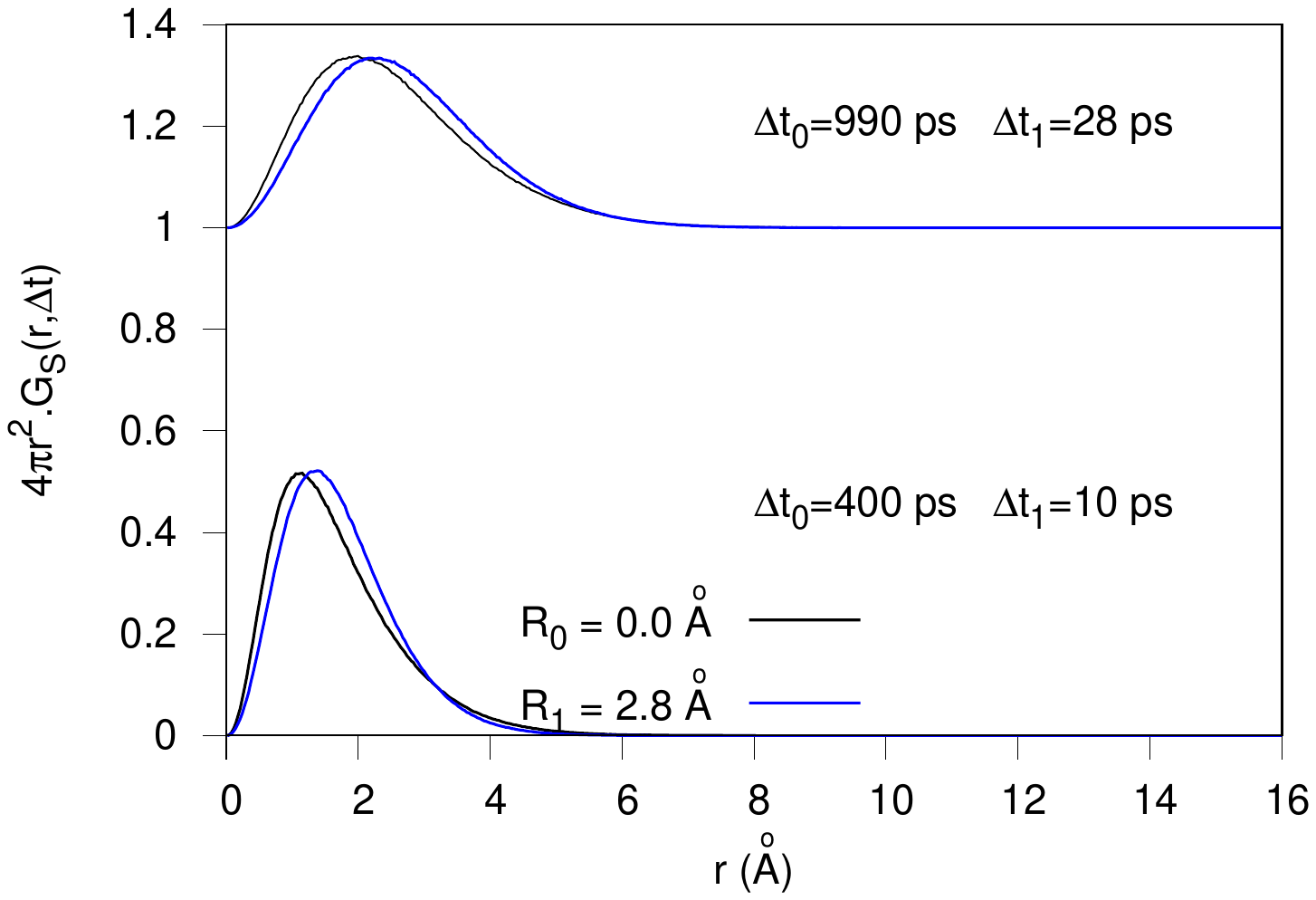}

\caption{(color online)  Self part of the Van Hove correlation function $G_{s}(r,\Delta t)$ for $R=2.8$ \AA\  compared with the original potential self Van Hove function ($R=0$ \AA). We observe that the self part of the dynamics is slightly slower than the distinct part (i.e. the structural dynamics displayed in Figure \ref{fig4}) when $R$ increases. We interpret that change as an effect of the decrease of the cooperative motions (or time fluctuations) when the time coarse graining increases. Due to the decrease of cooperative motions, the average displacement of a molecule decreases.
}
\label{fig5}
\end{figure}

\begin{figure}
\centering
\includegraphics[height=7.5 cm]{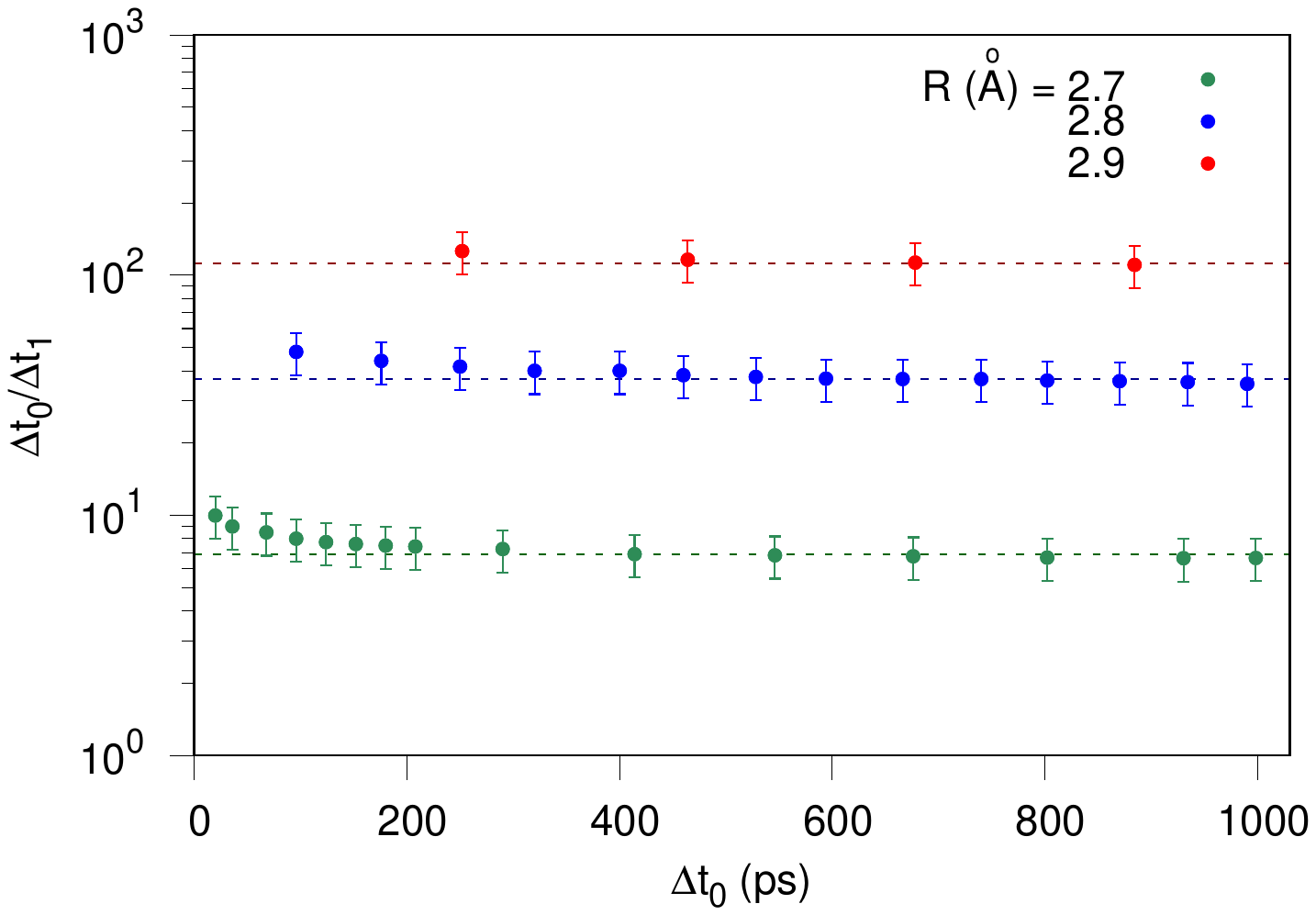}

\caption{(color online) {\color{black}Time rescaling ratio for the self part of the Van Hove correlation function, $\Delta t_{0}/\Delta t_{1}$ versus $\Delta t_{0}$,   where $G_{s}(r,\Delta t_{1})$ fits $G_{s}(r,\Delta t_{0})$, for (a) $R=2.7$ \AA, (b) $R=2.8$\AA, (c) $R=2.9$ \AA.}}
\label{fig5b}
\end{figure}

Let's turn now our attention to the self part of the Van Hove correlation function $G_{s}(r,\Delta t)$ in Figure \ref{fig5} {\color{black} and corresponding time ratios in Figure \ref{fig5b}.}
This correlation function measures the statistical displacement of molecules.
We observe that the self part of the dynamics is slightly changed in the model. We interpret that change as an effect of the decrease of the cooperative motions (or time fluctuations) when the time coarse graining increases. Due to the decrease of cooperative motions, the average displacement of a molecule decreases.\\

\begin{figure}
\centering
\includegraphics[height=7.5 cm]{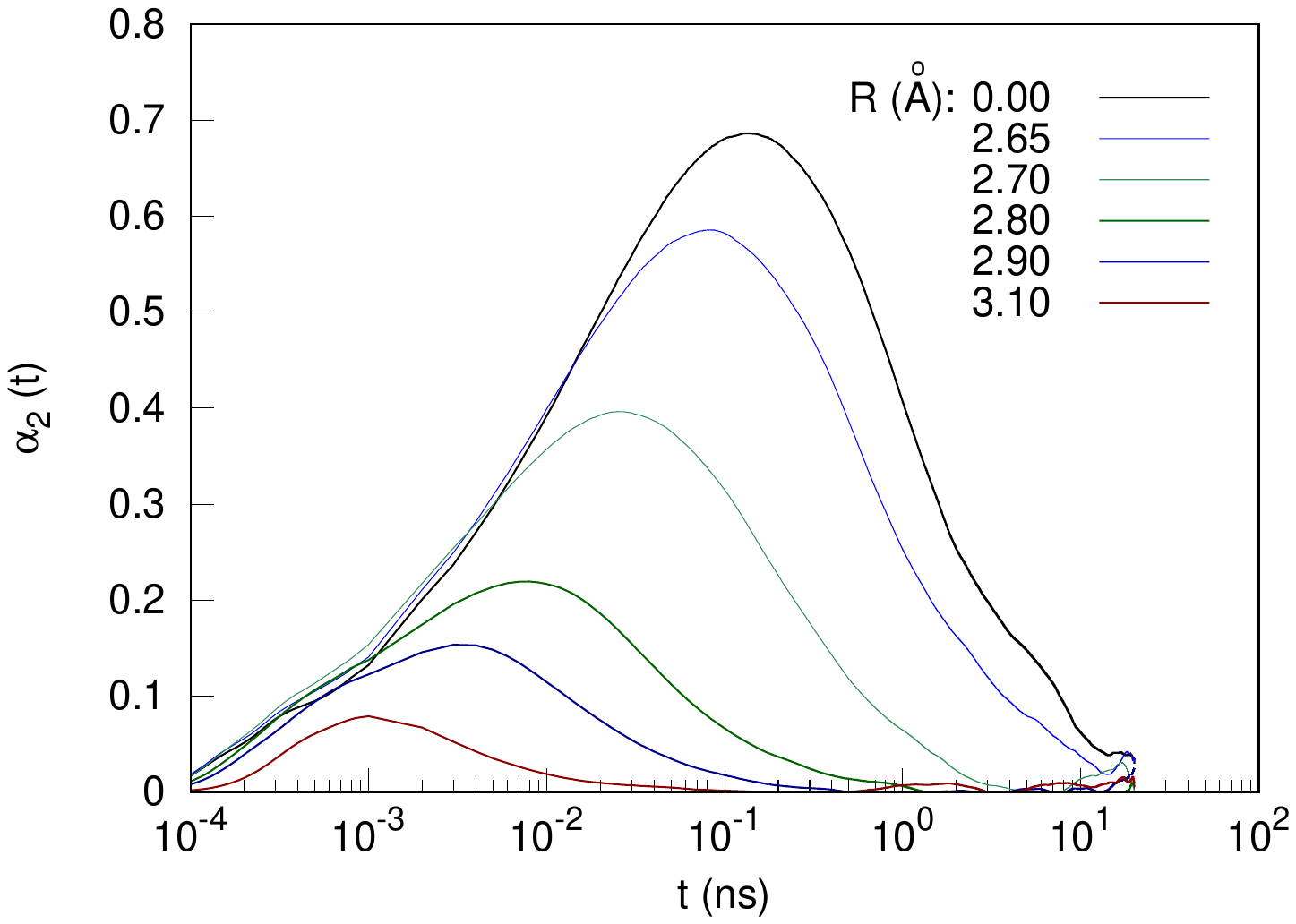}

\caption{(color online)  Non Gaussian parameter $\alpha_{2}(t)$ for various cutoff radii $R$. As $R$ increases (that is the time coarse grain increases). We observe a decrease of the Non Gaussian parameter. This decrease is expected due to the decrease of the size of correlated motions when the time coarse graining increases.}
\label{fig6}
\end{figure}

Supercooled liquids  {\color{black} are subject to} spontaneous cooperative motions (called dynamic heterogeneities)\cite{dh0} that we will now investigate using the Non Gaussian parameter (NGP) $\alpha_{2}(t)$.
 Figure \ref{fig6} compares the behaviors of the Non Gaussian parameters $\alpha_{2}(t)$ obtained for different coarse grain models and with the original potential. 
 We observe in the Figure a decrease of $\alpha_{2}(t)$ and of its characteristic time $t^{*}$ when $R$ increases. 
 
 {\color{black} We interpret this result as follows. When} the cutoff radius $R$ is made larger, the coarse grain time step increases, leading to a coarse grained value of the mobility fluctuations. The mobility fluctuations therefore decrease leading to a decrease of the NGP that essentially measures large deviations of these fluctuations.  This can be seen as the expected renormalization effect leading to the decrease of the correlation $\zeta(t)$. The renormalized value $\zeta(t)/\Delta t$ decreases as $\Delta t$ increases.\\

\begin{figure}
\centering
\includegraphics[height=7. cm]{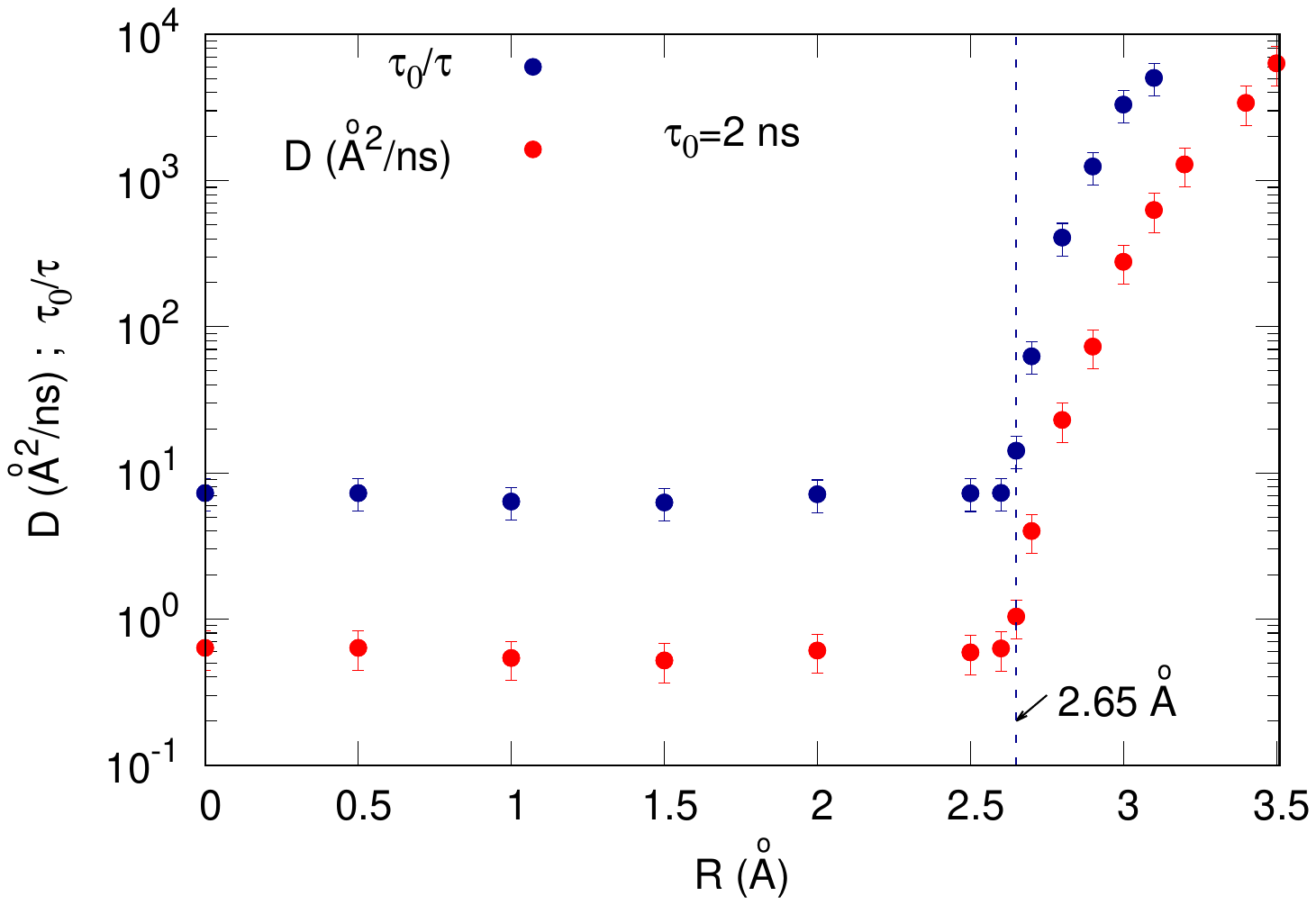}

\caption{(color online)  Diffusion coefficient $D$ and inverse of the $\alpha$ relaxation time ($1/{\tau}_{\alpha}$) versus the cutoff radius $R$. The two parameters are constant below $R=2.65$ \AA, then increase rapidly with $R$.}
\label{fig7}
\end{figure}

The main physics of the system has now been studied and we will now turn our attention to more practical data (see Figures \ref{fig7}, \ref{fig8}, \ref{fig9} and  \ref{fig10}).
In Figure \ref{fig7} we display the evolution of the diffusion coefficient $D$ and alpha relaxation time $\tau_{\alpha}$ with $R$.
As discussed before, an increase in the cutoff radius $R$ is equivalent to an increase of the time parameter $\Delta t$ in the {\color{black} GMF potential} $V^{gmf}(r,\Delta t)=-kT log(G_d(r,\Delta t)/\rho)$.

 As $R$ (or $\Delta t$) increases, the diffusion coefficient and alpha relaxation time stay constant, as long as $R<R_{T}=2.65$ \AA.
 Then above that threshold we observe a decrease of the $\alpha$ relaxation time $\tau_\alpha$ and an increase of the diffusion coefficient. Therefore the medium's viscosity decreases when $R$ increases. 
 
 This decrease of the relaxation time $\tau_\alpha$ could be interpreted as originating from an increase of an effective  temperature  (due to the potential modification), or as the expected consequence of the increase of the time step $\Delta t$. In our simulations the real temperature and therefore the kinetic energy do not change, but the decrease of the potential energy due to our modification of the potential short range wall could  induce a modification of the ratio potential energy/kinetic energy that would be equivalent to an increase of an effective temperature. 
{\color{black} The effect of the short range increase of the distinct Van Hove with $\Delta t$ leading to a decrease of the potential wall at short range could then be roughly interpreted as a possible origin of the time temperature superposition principle.}


\begin{figure}
\centering
\includegraphics[height=7.5 cm]{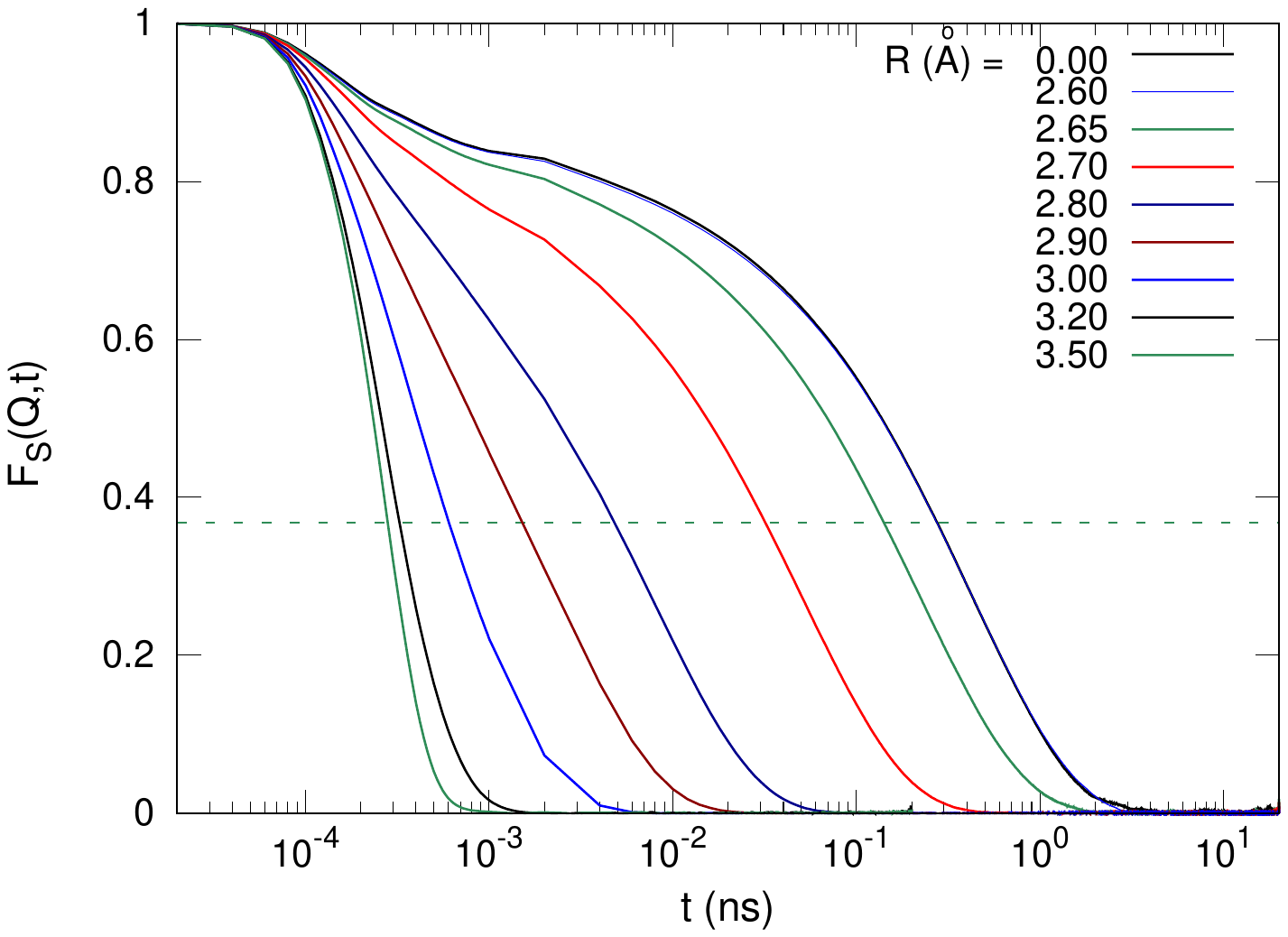}

\caption{(color online)  Incoherent scattering function $F_{S}(Q,t)$ versus time, for various cutoff radii $R$. As $R$ increases, the characteristic time $\tau_\alpha$  decreases, in agreement with an increase of time coarse graining.}
\label{fig8}
\end{figure}

In order to better understand the origin of the relaxation time decrease and diffusion increase, we will now study the correlation functions from which these data are obtained, that is the incoherent scattering function and the mean square displacement.
The incoherent scattering function $F_{S}(Q,t)$ (Figure \ref{fig8}) represents the Fourier transform of the local density autocorrelation function.
$F_{S}(Q,t)$ relates to the local dynamics of the liquid and gives us information on the $\alpha$ relaxation process. The $\alpha$ relaxation time $\tau_{\alpha}$ is obtained from $F_{S}(Q,t)$ with equation \eqref{e10} and is linked to the liquid viscosity.

Figure \ref{fig8} displays the time evolution of $F_{S}(Q,t)$ for various cutoff values $R$ or equivalently time steps $\Delta t$.
As long as $R$ is below the threshold the incoherent scattering function is not affected by the coarse graining.
While above the threshold ($R>2.65 $\AA), the  plateau regime first decreases, then disappears.
We observe the same behavior for the mean square displacement (Figures \ref{fig9} and \ref{fig10}). As $R$ increases, the plateau first decreases, then disappears. \\

We interpret the decrease of the plateau as originating from the increase in the time step $\Delta t$ that is the size of time coarse graining.
 Figure \ref{fig10} shows the time rescaled MSD, the rescaling being chosen to lead to the same diffusion coefficient, that is the same MSD for large times. 
The Figure shows that the cage escaping process appears at shorter time as $R$ increases, leading to a shortening of the plateau which is a characteristic of supercooled media. Eventually when  $R>3.1$ \AA, the plateau disappears and the MSD is characteristic of a simple liquid.
As the time coarse grain increases the size of the plateau decreases, then disappears when the coarse grain time reaches the diffusive time scale.

\begin{figure}
\centering
\includegraphics[height=7.5 cm]{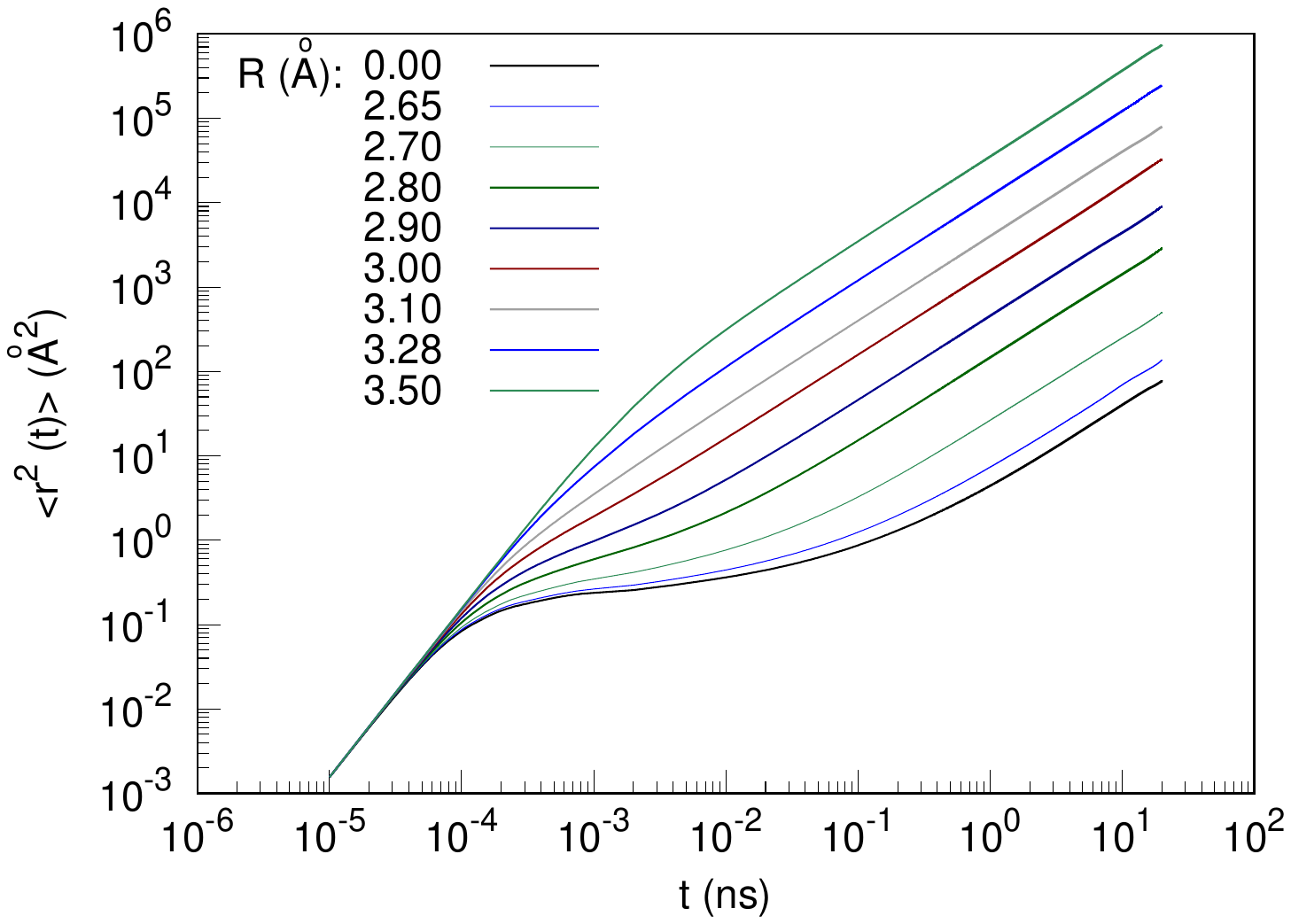}

\caption{(color online)  Mean square displacement (MSD) $<r^{2}(t)>$ time evolution for various cutoff radii $R$. The short time mean square displacement doesn't change. Due to its ballistic nature, it depends only on the constant temperature of the liquid.
However for larger time scales the MSD evolution with $R$ is similar to the effect of an increase in temperature (an effective temperature here, that would be due to the cage softening), or to a shift in time due to time coarse graining.}
\label{fig9}
\end{figure}

\begin{figure}
\centering
\includegraphics[height=7.5 cm]{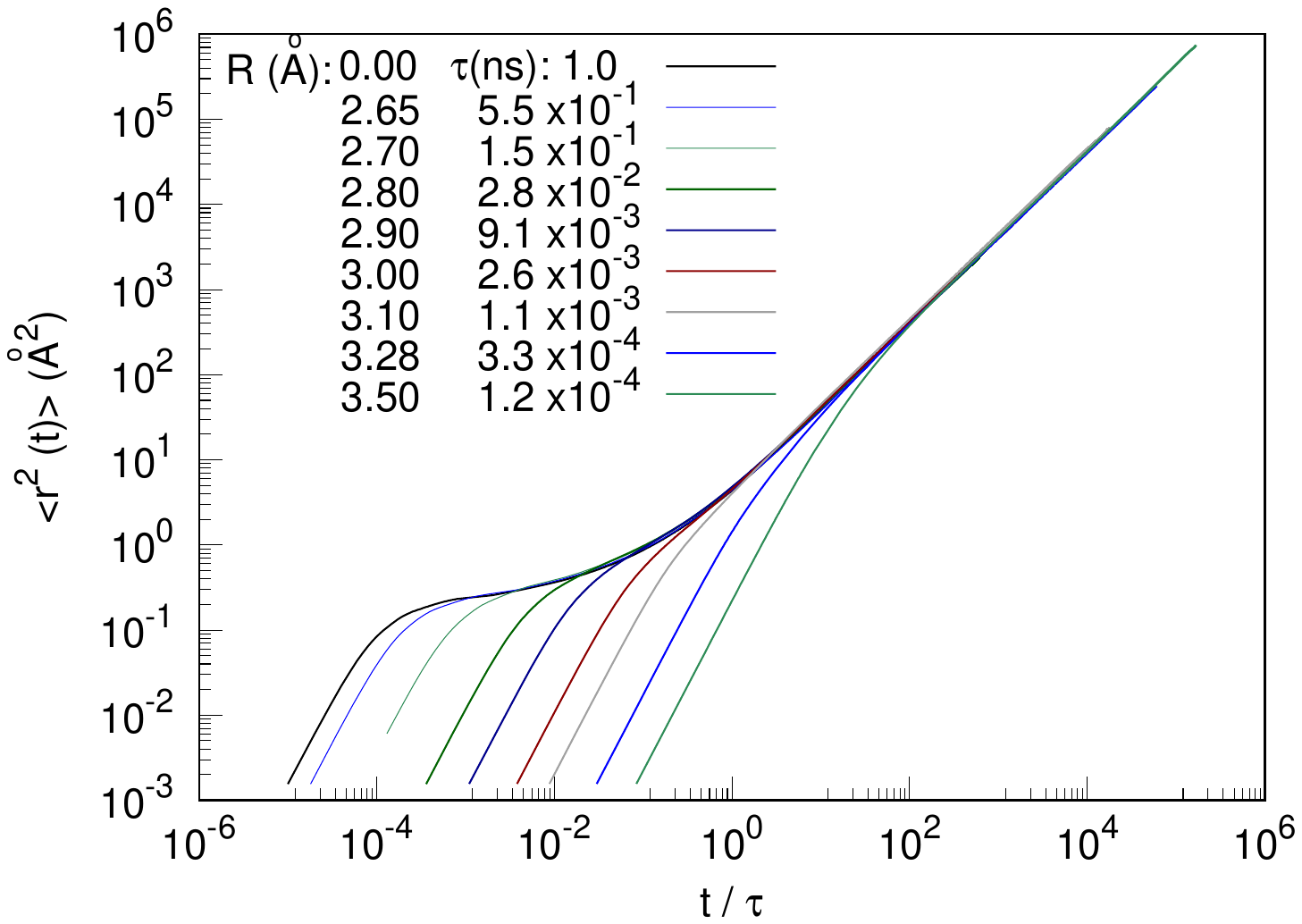}

\caption{(color online)  Time rescaled Mean square displacement $<r^{2}(t)>$ for various cutoff radii $R$. {\color{black} The rescaling time $\tau$ evolves with the coarse graining parameter $R$ as expected for a coarse graining in time.}}
\label{fig10}
\end{figure}

{\color{black} Finally, we found that the Van Hove distinct was well reproduced with a time step $\Delta t_{1}$ that was increasing with $R$ as expected for a time coarse graining controlled by the cutoff $R$ .
The Van Hove self correlation function however was reproduced with a relatively smaller time step, and the Non Gaussian parameter decreased with $R$.  
We interpret these differences as follows. Due to time coarse graining, the cooperative motions called 'dynamic heterogeneity' decrease as they are coarse grained.
This results in a decrease of the Non Gaussian parameter that measures the heterogeneity and a decrease of the average self motion of molecules, as most mobile molecules decrease in number. 

We have to notice also as a drawback that the method we describe doesn't take into account modifications of the interactions directions during the time step.
It may then be limited to time steps and systems for which it is not important as in supercooled liquids and soft matter due to cage effect, or the resulting uncertainty must be taken into account.}

\section{Conclusion}

In this work we investigated the possibility of time coarse graining molecular dynamics simulations. 
Similarly to Boltzmann inversion method in spatial coarse graining, which begins with a free energy called potential of mean force $V^{mf}(r)=-kT log (g(r))$, we tested the effect of a generalized potential of mean force procedure that uses the distinct part of the Van Hove correlation function with a characteristic time different from zero $V^{gmf}(r,\Delta t)=-kT log (G_{d}(r,\Delta t)/\rho)$. 
We found that the short range part of $V^{gmf}(r,\Delta t)$ follows {\color{black} a quadratic law.
We then defined our effective potential with a quadratic law at short range, replacing the short range wall of the Lennard-Jones potential with a quadratic law and keeping the original Lennard-Jones potential at larger ranges. Our effective potential is then defined by a single cutoff parameter $R$ that separates the quadratic law from the original potential, with the conditions that the potential and force are continuous at that point.
We then tested that very simple time coarse grained modeling} and found that it actually leads approximately to the same dynamics as the original potential with a much larger time step. 

Interestingly enough, the quadratic smoothing of the short range wall when time is coarse grained {\color{black}could well be related to} the time temperature superposition principle, and the short range peak\cite{dh0} that appears for most mobile molecules in $G_{d}(r,\Delta t)/\rho$ at $\Delta t=t^{*}$ in supercooled liquids, will lead to an attractive potential function for mobile molecules that could be interpreted as a facilitation mechanism. {\color{black}Work is in progress to better understand these possible physical perspectives that we emphasize are still speculative.}

Finally, a simple modification of the potential, namely replacing the short range wall with a smooth quadratic law, could lead to a shift in the time step resulting in a similar dynamics as the original potential function but with a much larger time step.   The method while used here for a supercooled liquid is expected to hold in various systems.

}









\end{document}